\documentstyle[preprint,aps,epsf]{revtex}
\newcommand{\beq}{\begin{equation}}
\newcommand{\eeq}{\end{equation}}
\newcommand{\beqar}{\begin{eqnarray}}
\newcommand{\eeqar}{\end{eqnarray}}
\newcommand{\ds}{\displaystyle}
\begin{document}
\tightenlines
\draft

\title  {
 Transverse momentum dependence of directed particle flow 
 at 160{\it A}~GeV
         }
\author {
E.~E.~Zabrodin$^{1,2}$, C.~Fuchs,$^{1}$, L.~V.~Bravina,$^{1,2}$ 
and Amand~Faessler$^{1}$ \\
  }
\address{
 $^1$Institute for Theoretical Physics, University of T\"ubingen,\\
 Auf der Morgenstelle 14, D-72076 T\"ubingen, Germany
         }
\address{
 $^2$Institute for Nuclear Physics, Moscow State University,
 RU-119899 Moscow, Russia
         }

\maketitle

\begin{abstract}
The transverse momentum ($p_t$) dependence of hadron flow at SPS 
energies is studied. In particular, the nucleon and pion flow in 
S+S and Pb+Pb collisions at 160{\it A} GeV is investigated. 
For simulations the microscopic quark-gluon string model (QGSM) 
is applied. It is found that the directed flow of pions
$v_1(y, \Delta p_t)$ changes sign from a negative slope in the 
low-$p_t$ region to a positive slope at $p_t \geq 0.6$~GeV/$c$ as 
recently also observed experimentally. The change of the flow
behaviour can be explained by early emission times
for high-$p_t$ pions. We further found that a substantial 
amount of high-$p_t$ pions are produced in the very first 
primary NN collisions at the surface region of the touching 
nuclei. Thus, at SPS energies high-$p_t$ nucleons seem
to be a better probe for the hot and dense early phase of 
nuclear collisions than high-$p_t$ pions. Both, in the light and 
in the heavy system the pion directed flow $v_1(p_t, \Delta y)$
exhibits large negative values when the transverse momentum
approaches zero, as also seen experimentally in Pb+Pb collisions.
It is found that this effect is caused by nuclear shadowing.
The proton flow, in contrary, shows the typical linear increase
with rising $p_t$.

\end{abstract}
\pacs{PACS numbers: 25.75.-q, 25.75.Ld, 24.10.Lx, 24.10.Jv}

\section{Introduction}
\label{sec1}

The collective flow of hadrons in ultrarelativistic heavy-ion 
collisions is a very useful signal to probe the evolution of hot and 
dense nuclear matter from the onset of its formation 
\cite{StGr86,ReRi97,HWW99,Dan99,Olli92,Cser94,lca}. 
Since the development of flow is closely related to the equation of
state (EOS) of nuclear matter, the investigation of the flow can shed
light on the transition to a new phase of matter, the so-called
quark-gluon plasma (QGP), and its subsequent hadronization 
\cite{Amprl91,HuSh95,RiGy96,Brprc94,Sol97,Sorplb97,Sorprl97,Sorprl99,
HeLe99,CsRo99,Ri95,Brac00,Bao99,KSH00,e877,na49,PoVoqm,wa98a,wa98b}.
If the transition from the QGP to hadronic phase is of first order, 
the vanishing of the pressure gradients in the mixed phase leads to
the so-called softening of the EOS \cite{HuSh95,RiGy96}. The latter 
should be distinctly seen in the behaviour of the excitation function
of the collective flow.

At present, the Fourier
expansion technique is widely employed to study collective flow
phenomena \cite{VoZh96,Vol97,PoVo98}. Namely, the invariant 
distribution $\ds E \frac{d^3 N}{d^3 p}$ is presented as 
\beq
\ds
E \frac{d^3 N}{d^3 p} = \frac{1}{\pi} \frac{d^2 N}{dp_t^2 dy} \left[ 
1 + 2 \sum_{n=1}^{\infty} v_n \cos(n\phi) \right] ,
\label{eq1}
\eeq
where $p_t$ and $y$ are the transverse momentum and rapidity, and 
$\phi$ is the azimuthal angle between the momentum of the particle 
and the reaction plane. The first two Fourier coefficients in
Eq.(\ref{eq1}), $v_1$ and $v_2$, are dubbed directed and elliptic 
flow, respectively. Since both types of anisotropic flow depend
on rapidity and transverse momentum, one is able to study double
differential distributions
\beq
\ds
v_n(p_t,\Delta y) = \int \limits_{y_1}^{y_2} \cos(n\phi) 
\frac{d^2 N}{dp_t^2 dy} dy \left/ \int \limits_{y_1}^{y_2} 
\frac{d^2 N}{dp_t^2 dy} dy \right.
\label{eq2}
\eeq
and
\beq
\ds
v_n(y, \Delta p_t) = \int \limits_{p_t^{(1)}}^{p_t^{(2)}} 
\cos(n\phi) \frac{d^2 N}{dp_t^2 dy} dp_t^2 \left/
\int \limits_{p_t^{(1)}}^{p_t^{(2)}} 
\frac{d^2 N}{dp_t^2 dy} dp_t^2 \right. \ .
\label{eq3}
\eeq

Model calculations show that elliptic flow is built up at the
early phase of nuclear collisions \cite{Sorplb97,KSH00}, whereas
directed flow develops until the late stage of the reaction
\cite{LPX99}. But it is well known that the particles with high
transverse momentum are emitted at the onset of the collective
expansion, i.e., their directed flow can carry information about
the EOS of the dense nuclear phase. Our first goal is to check the
emission times of high-$p_t$ hadrons and to study the transverse
momentum dependence of directed flow in heavy-ion collisions at
SPS (160{\it A\/}~GeV) energies.

Apparently, one would expect that the directed flow drops to zero
when the transverse momentum decreases. For protons such a 
behaviour has been observed at SIS energies ($\leq 1$AGeV) 
\cite{andronic}. Experimental results on pion and proton directed 
flow at both AGS \cite{e877} and SPS \cite{na49} energies, 
show a qualitatively different picture: $v_1(p_t)$ is positive at 
high $p_t$ and slightly but negative at low transverse momenta, i.e., 
it approaches zero from the negative side. One of the possible 
explanations of such a peculiar behaviour has been proposed in 
\cite{Vol97} within the framework of a thermal model. Here the 
interplay of the radial expansion of a thermalized source and the 
directed flow has been discussed. It was shown that the $v_1(p_t)$ of 
protons became negative at small values of the transverse momentum 
provided the transverse expansion velocity of a thermalized source was 
$\beta \cong 0.55c$ or higher. However, the significance of, e.g.,
an initial anisotropy of the geometrical configuration in non-central 
collision for the development of anisotropic flow is not completely 
understood at energies above 10{\it A}~GeV. The aim of the present 
paper is also to elaborate the role of non-dynamic, i.e. geometrical 
effects, which can cause the preferential emission of particles in the 
direction opposite to that of the normal flow (the so-called antiflow) 
at small $p_t$. For this purpose the quark-gluon string model (QGSM) 
\cite{qgsm} is chosen. 

The paper is organised as follows. A brief description of the 
microscopic model is given in Sect.~\ref{sec2}. Rapidity and 
transverse momentum dependences of frozen nucleons and pions,
calculated in Pb+Pb central collisions at 160{\it A\/}~GeV, are
also discussed. Section~\ref{sec3} presents a systematic study of the 
directed pion and nucleon flow at SPS energies as a function of 
$p_t$ and $y$. A comparison with experimental data 
is performed as far as those are available. The directed
flow of nucleons, which is developed alongside of the normal flow,
grows with rising transverse momentum, while the directed flow of
pions changes its orientation from antiflow at low $p_t$'s to
normal flow at high transverse momenta. In order to compare to 
the predictions of the thermal approach \cite{Vol97} as well, the 
calculations for the proton flow are fitted by an expanding 
thermalized source. Finally, conclusions are drawn in 
Sect.~\ref{sec4}.

\section{Features of particle production and freeze-out in the 
model}
\label{sec2}

The QGSM, which treats the elementary 
hadronic interactions on the basis of Gribov-Regge theory, 
is based on the $1/N_c$ (where $N_c$ is the number of quark 
colours or flavours) topological expansion of the amplitude for 
processes in quantum chromodynamics and string phenomenology of 
particle production in inelastic binary collisions of hadrons. 
The model incorporates the production of particles via string
excitation and subsequent fragmentation, as well as the formation of
resonances and hadron rescattering. As independent degrees of freedom 
the QGSM includes octet and nonet vector and pseudoscalar mesons, and 
octet and decuplet baryons, and their antiparticles.
The model simplifies the in-medium effects and focuses mainly on the 
development of an intranuclear cascade. Further details on the QGSM 
can be found elsewhere \cite{qgsm}.

For the simulations at SPS energies, $E_{lab}=160${\it A} GeV, light 
$^{32}$S+$^{32}$S and heavy $^{208}$Pb+$^{208}$Pb symmetric systems
have been chosen. According to the QGSM predictions \cite{frprc99}
the mean number of interactions per hadron, $\langle N^h_{int}
\rangle $, equals 2 even for central sulphur-sulphur collisions
and 9 for central lead-lead collisions at 160{\it A} GeV. Due to a 
significant increase of $\langle N^h_{int} \rangle $ with rising mass 
number, one can study the role of the intranuclear cascade on the 
formation of transverse collective flow. Although the light S+S 
system, or rather a part of it, cannot be treated as a thermalized 
source, the formation of thermally equilibrated matter in Pb+Pb 
collisions is not ruled out. The system expands until all interactions
and decays in the reaction have ceased. This stage corresponds to the 
conditions of the thermal freeze-out. Note also that the system of 
final particles in the course of model simulations may be well 
approximated by a core and a halo structure. The halo contains
frozen particles already decoupled from the main system, and the
core consists of hadrons intensively interacting, both elastically
and inelastically, with each other. In other words, there is no 
sharp freeze-out picture in the microscopic model like the QGSM
\cite{frprc99} or the relativistic quantum molecular dynamics
(RQMD) model \cite{Sor96}, in contrast to macroscopic hydrodynamic
models (see, e.g., \cite{Mag99,BGG99} and references therein).  

\begin{figure}[htp]
\centerline{\epsfysize=16cm \epsfbox{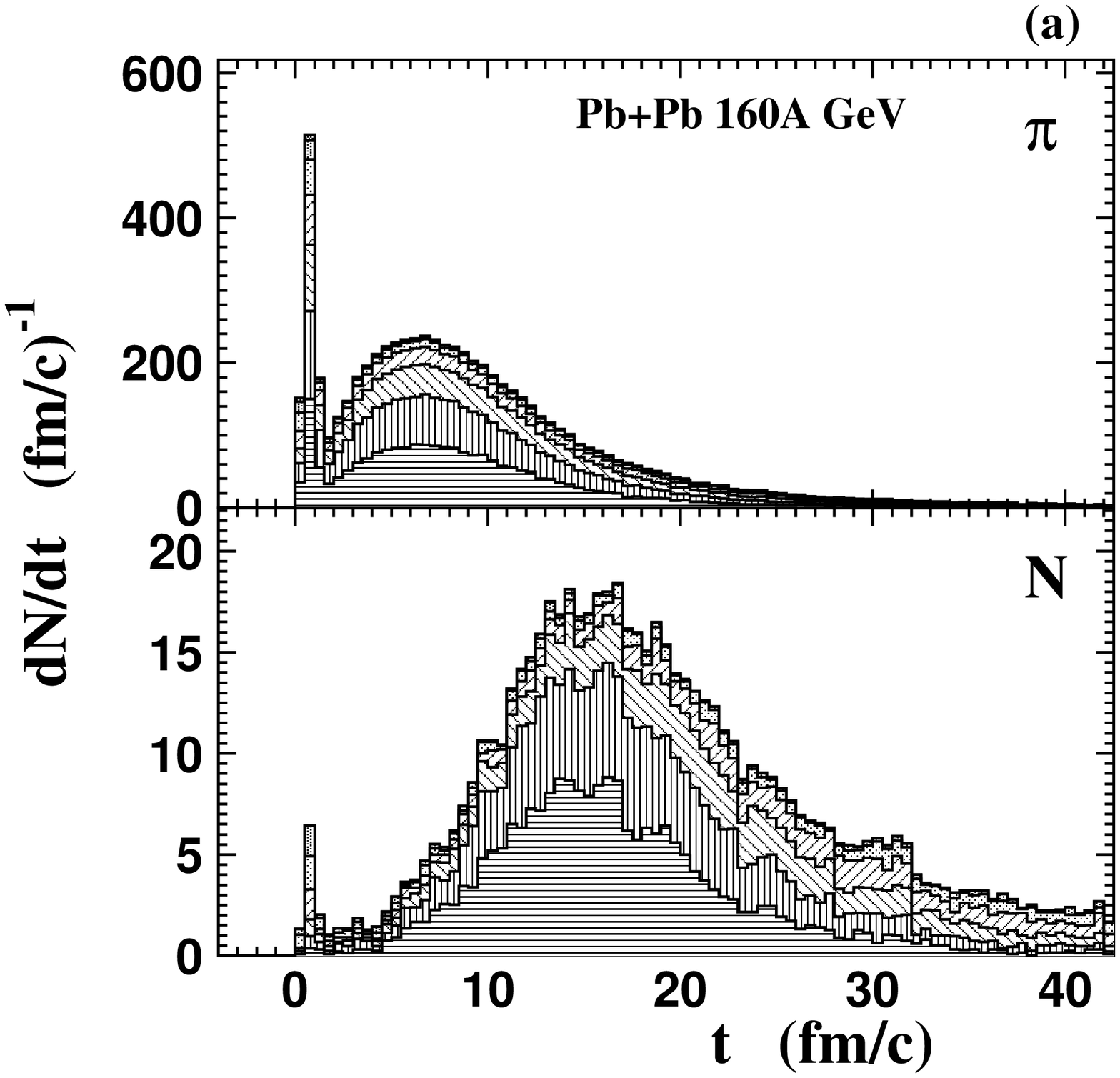}}
\caption{
(a) Evolution of number of frozen pions (upper panel) and nucleons
(lower panel) with time $t$ in central Pb+Pb collisions
at 160{\it A}~GeV. Hatched areas correspond to the
central-mass-frame rapidity intervals
$|y_{c.m.}| < 0.5,\, 1.0,\, \ldots\, 3.5\ .$ \\
(b) Transverse mass distribution $d^2 N / m_T d m_T d t / A$ of the
final-state nucleons (upper panel) and pions (lower panel) according
to their emission time $t$. Contour plots correspond to densities
$d^2 N / m_T d m_T d t / A = 0.003,\ 0.01,\ 0.03,\ 0.1,\ 0.25,\ 1.0,
\ 3.0,\ 10.0$ particles per fm$^2/c$.
}
\centerline{\epsfysize=16cm \epsfbox{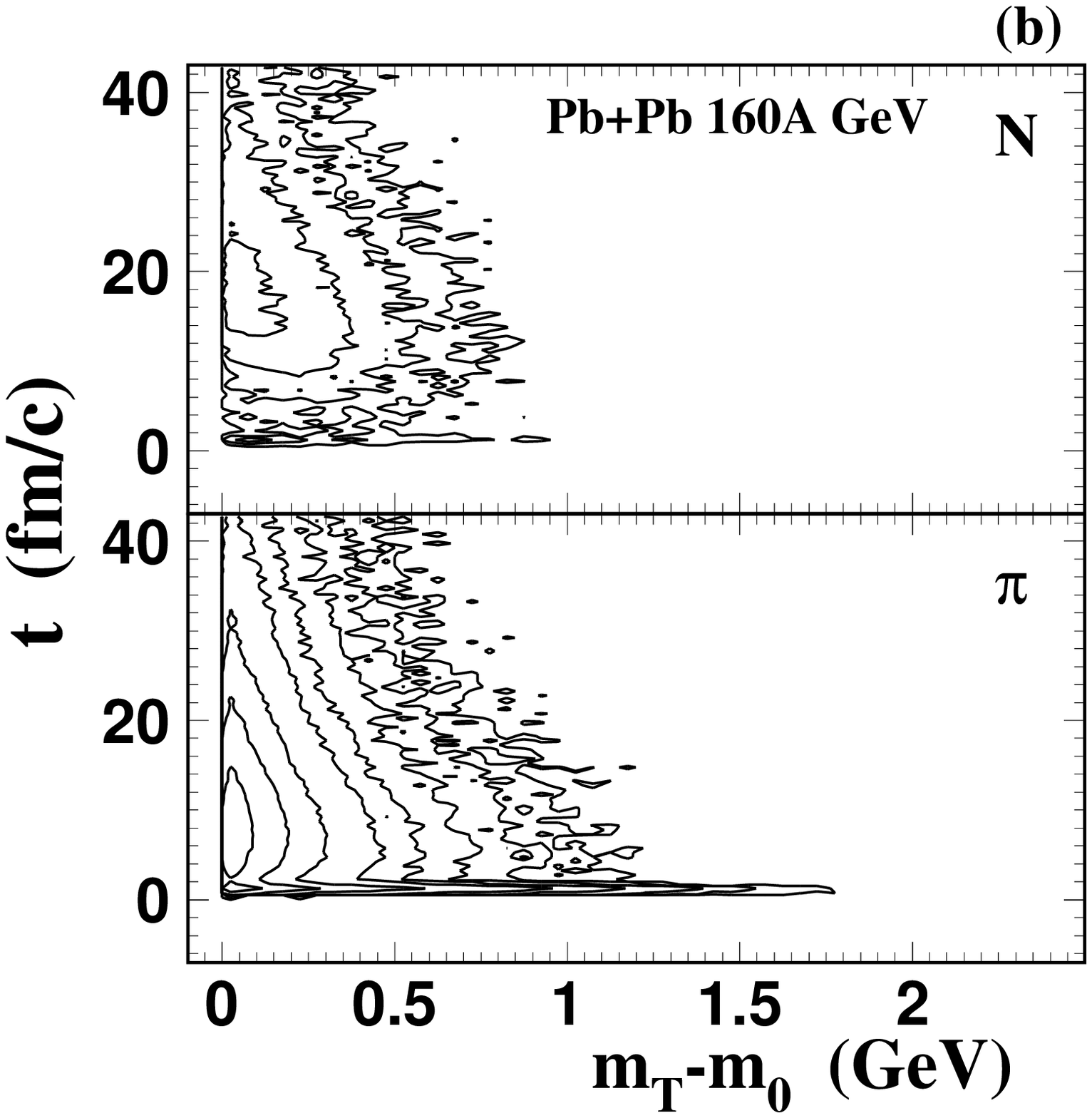}}
\label{fig1}
\end{figure}

The evolution of the number of frozen particles with time $t$ is 
shown in Fig.~\ref{fig1}(a) for pions and nucleons in Pb+Pb central 
collisions. The contour plots (dashed areas) correspond to the 
$d N/ dt$ distribution in different rapidity intervals. One can see
that the pionic distributions peak at $t \approx 8$ fm/$c$, and the
nucleon distributions reach their maxima later, at $t \approx 15$
fm/$c$. It is interesting that the positions of the maxima on the 
time scale are not shifted when the rapidity range is enlarged. 
Many hadrons with high rapidity, especially pions, 
are emitted from the very beginning of the nuclear collision. These 
particles usually have rather high transverse momentum as well. 
To illustrate this idea, the time evolution of the transverse mass
distributions of nucleons and pions at the freeze-out is shown
in Fig.~\ref{fig1}(b). We see that nucleons with maximal
transverse momenta in lead-lead collisions are coming either
from the very beginning of the reaction or from intermediate times 
with a maximum at $t \approx 13$ fm/$c$. In contrast to nucleons
pions with highest $p_t$ are produced in inelastic primary NN 
collisions in heavy-ion reactions, while soft particles are 
emitted during the whole evolution time. This is a general trend 
in the production of soft and hard particles in relativistic
heavy-ion collisions. Since the excitation function of the transverse 
particle flow is very sensitive to pressure gradients in the
system, the directed flow of both pions and 
nucleons might change its behaviour with increasing $p_t$. 
Obviously, pions coming from primary NN collisions cannot carry
information about properties of hot and dense nuclear matter, nor
about the relaxation process. These particles are just produced 
in the surface regions of the touching nuclei. 
The admixture of such pions will
severely distort the spectrum of pions stemming from the (nearly)
thermalized source. In contrast, the fraction of high-$p_t$ 
nucleons emitted promptly after the primary collisions is relatively
small compared to total number of high-$p_t$ nucleons. Therefore,
the spectrum of nucleons with large transverse momentum in heavy-ion
collisions at SPS energies might be even more useful to 
study the early stage of the fireball evolution than that of pions.

\section{Directed flow of hadrons}
\label{sec3}

\subsection{Comparison with experimental data}
\label{subsec3a}

\begin{figure}[htp]
\centerline{\epsfysize=16cm \epsfbox{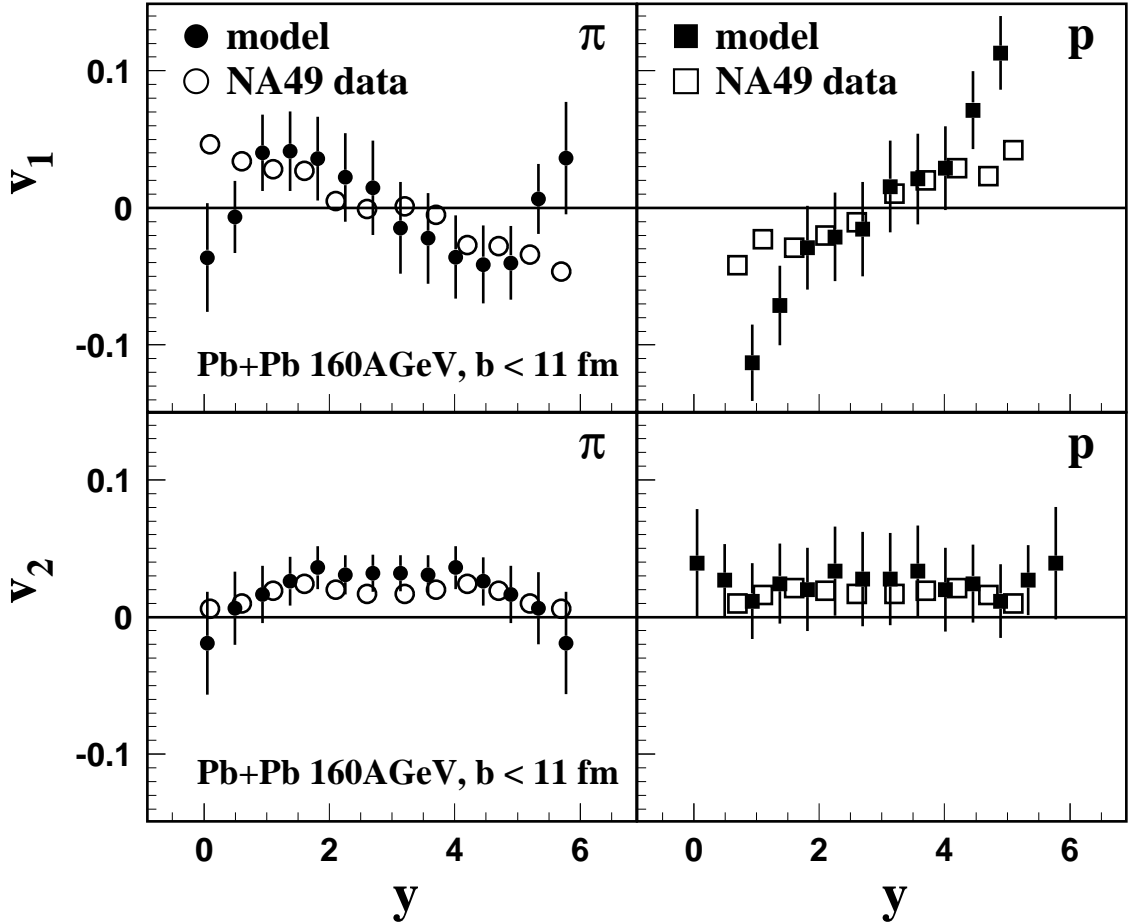}}
\caption{
Directed flow (upper row) and elliptic flow (lower row) for pions
(left panels) and protons (right panels) in minimum bias Pb+Pb
collisions at SPS energies. Solid symbols indicate microscopic
calculations, open symbols show the experimental data taken from
\protect\cite{PoVoqm}.
}
\label{fig2}
\end{figure}

First, directed flow and elliptic flow of pions and protons,
calculated for Pb+Pb minimum bias events with the maximum impact
parameter $b_{max} = 11$~fm at SPS energies, are compared in 
Fig.~\ref{fig2} with the experimental data of the NA49 Collaboration
\cite{PoVoqm}. Both for pions and for protons the agreement between 
the microscopic calculations and the experimental data, as well as
with the relativistic quantum molecular dynamics (RQMD) model 
results (see Fig.~1 of \cite{LPX99}), is quite reasonable. 
The directed flow of protons has a positive slope in the midrapidity 
range, which becomes steeper as the rapidity window is shifted 
towards projectile/target rapidity. Also, the pionic directed flow, 
which has a characteristic negative slope
of $v_1(y)$ in the range $1 \leq y \leq 5$, drops to zero and 
even becomes positive at $y \approx y_{max}$. This behaviour can be
understood, provided the fast particles are formed on the leading
quarks (mesons) and diquarks (baryons) at the early times of the
collision. The number of secondary interactions per particle with
high rapidity is small, thus, the flow in the fragmentation 
regions is basically determined by the initial geometry of the
system \cite{LPX99}. The centrality dependence of anisotropic flow
in the QGSM has been studied in \cite{Br95,BZFFplb99,BZFFprc00}. 
However, as was discussed in Sect.~\ref{sec2}, the directed flow of 
soft particles should differ from that of hard particles. Therefore, 
the $v_1(y, \Delta p_t)$ distribution given by Eq.~(\ref{eq3}) is 
used to study the transverse momentum dependence of directed flow. 

\subsection{$v_1(y)$ in $p_t$ intervals}
\label{subsec3b}

\begin{figure}[htp]
\centerline{\epsfysize=16cm \epsfbox{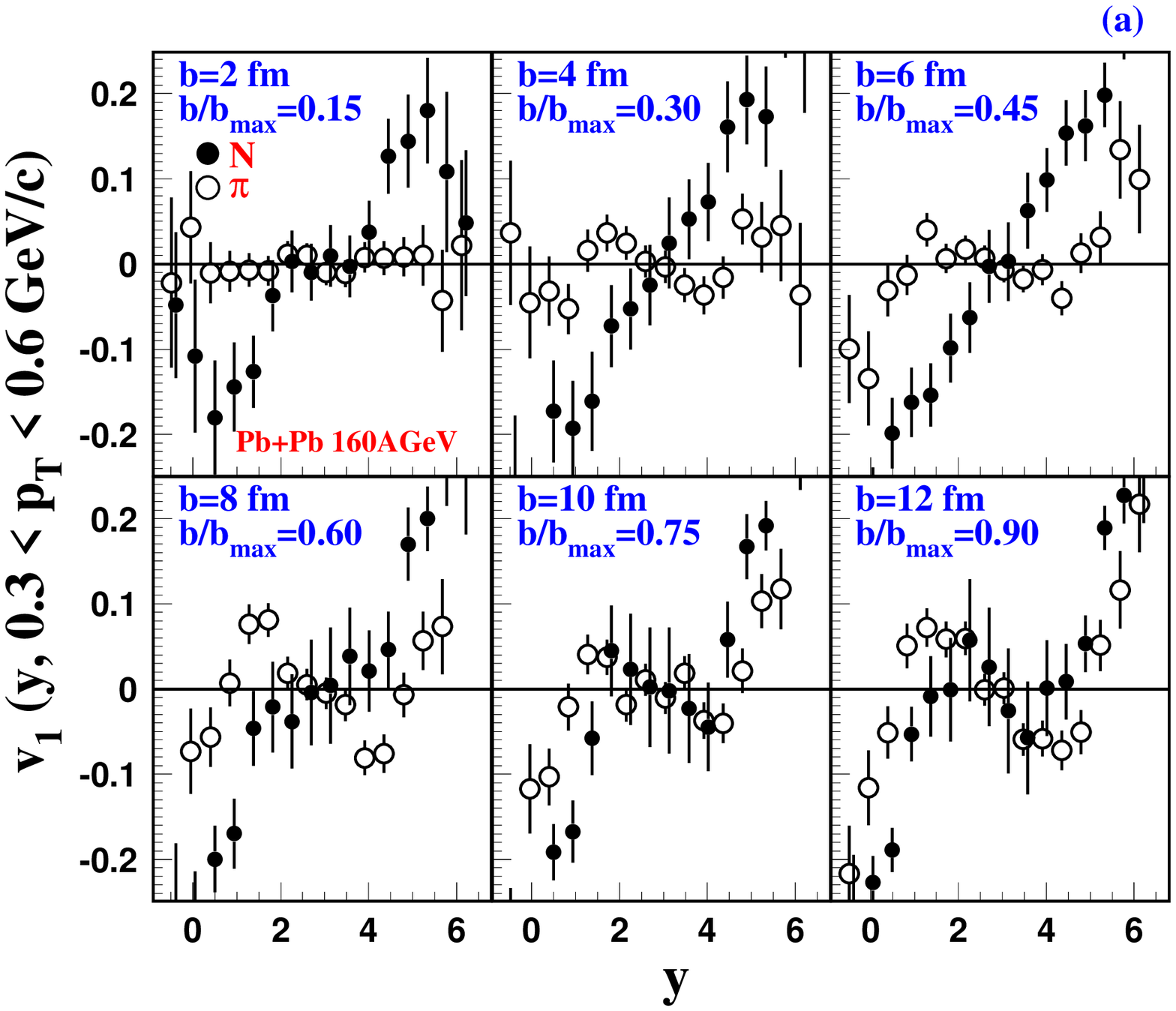}}
\caption{
(a) Centrality dependence of directed flow, $v_1(y,\Delta p_t)$, 
of nucleons (solid circles) and pions (open circles) in the
transverse momentum interval $0.3 < p_t < 0.6$ GeV/$c$ in Pb+Pb
collisions at 160{\it A\/}~GeV.\\
(b) The same as (a) but for  $0.6 < p_t < 0.9$ GeV/$c$.
}
\label{fig3}
\centerline{\epsfysize=16cm \epsfbox{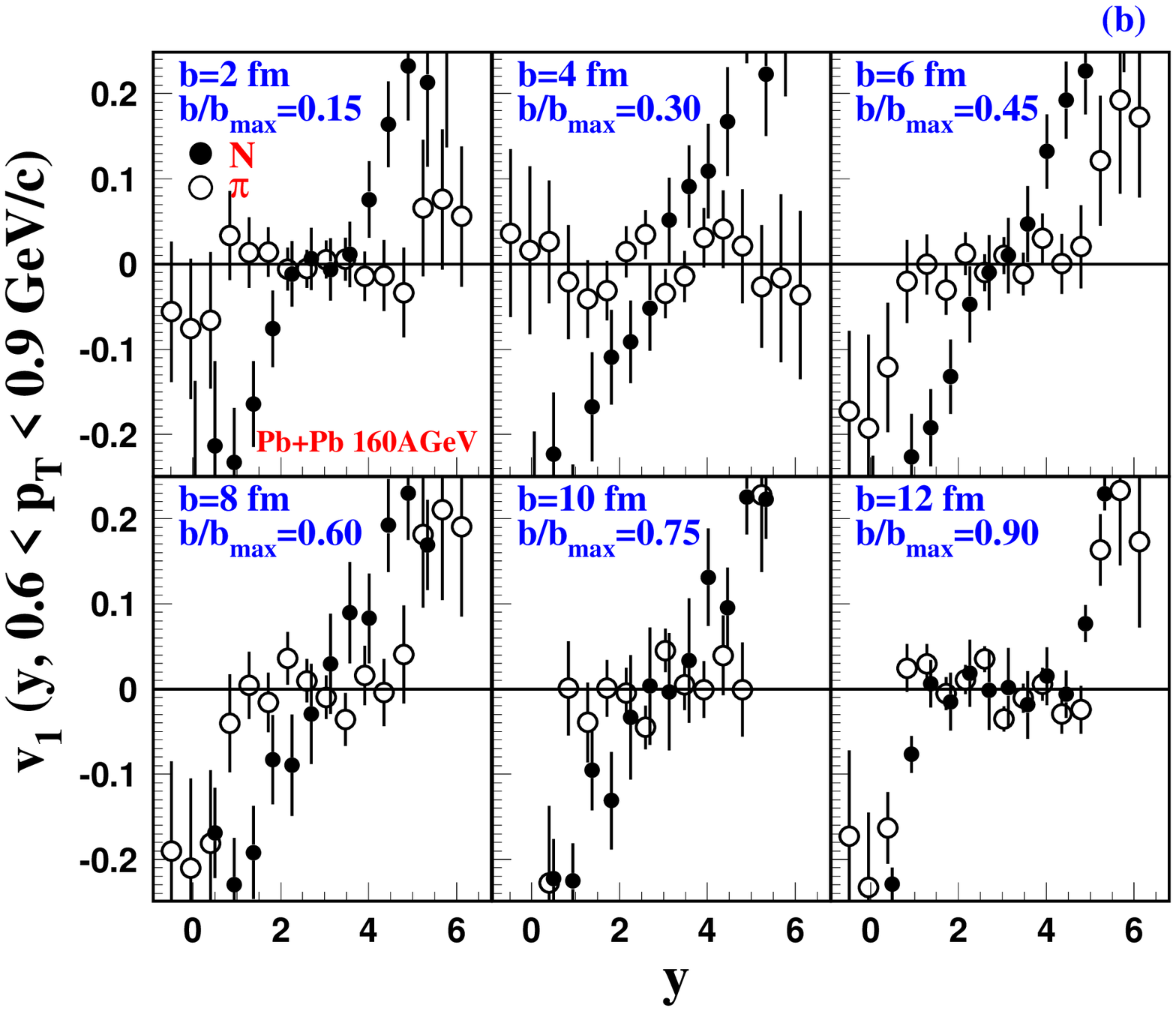}}
\end{figure}

Figure \ref{fig3} depicts the directed flow of nucleons and pions in 
two $p_t$ intervals, $0.3 < p_t < 0.6$ GeV/$c$ and 
$0.6 < p_t < 0.9$ GeV/$c$, for Pb+Pb collisions with different
centrality. The maximum impact parameter for a symmetric system is 
$b_{max} = 2\, R_A$. The value of the reduced impact parameter
$\tilde{b}=b/b_{max}$ in the simulations varies from 0.15 (central 
collisions) up to 0.9 (most peripheral collisions).
At $p_t < 0.6$ GeV/$c$ the pionic flow exhibits the typical
antiflow in both, semicentral and peripheral, collisions. In the
same $p_t$ interval the nucleon flow increases as the reaction 
becomes more peripheral. But at $b \approx 8$ fm the flow becomes 
softer in the midrapidity range. In very peripheral collisions the
directed flow of nucleons shows an antiflow behaviour which is similar 
to that of the pionic directed flow. As was shown in 
\cite{Br95,BZFFplb99,BZFFprc00}, see also \cite{LPX99,Sn99}, such a 
transformation of the nucleon flow is explained merely by shadowing: 
It is well known that the presence of even a small amount of 
quark-gluon plasma leads to a softening of the equation of state, 
which results in a significant reduction of the directed flow. 
However, since the QGP is expected to be produced primarily in central 
heavy-ion collisions, the effect should be most pronounced in central
collisions. In contrast, shadowing causes the disappearance of
nucleon directed flow and the development of antiflow in the 
midrapidity region especially in semiperipheral and peripheral
collisions, as well as in light systems.

The behaviour of directed flow changes drastically in the 
transverse momentum range $0.6 < p_t < 0.9$ GeV/$c$, presented in 
Fig. \ref{fig3}(b). Although the nucleon directed flow decreases in 
the midrapidity range at $b \geq 10$ fm, its normal component still 
dominates over the antiflow counterpart. Moreover, even 
high-$p_t$ pions prefer the direction of normal flow, distinctly
seen in semiperipheral events with $4 \leq b \leq 6$ fm. 
This reflects again that hadrons with high transverse momenta
($p_t \geq 0.6$ GeV/$c$) are produced mainly at the early stage
of nuclear collisions, as was discussed in Sect.~\ref{sec2}, see 
Fig.~\ref{fig1}(b).

The transition of the directed flow of pions from antiflow to normal
flow with rising transverse momentum has recently been observed in
Au+Au collisions at 1 AGeV \cite{Wag00}. The effect is stronger in
peripheral collisions and at target rapidities. Likely, this can be 
explained by the effective shadowing caused by the spectator matter,
in accord with quantum molecular dynamics (QMD) transport 
calculations. A similar feature is seen at SPS at projectile 
rapidities (see below). However, there is a difference in 
correlations between early freeze-out
times and high transverse momentum of hadrons in heavy-ion collisions
at 1{\it A\/}~GeV and 160{\it A\/}~GeV: At 1{\it A\/}~GeV high-$p_t$ 
pions are emitted within the first 15-20~fm/$c$ of the 
reaction \cite{Bass93,LBB91,uma97} and can be used as a
 ``time clock'' for the reaction which probes the high 
density phase \cite{Wag00}. However, at SIS energies high-$p_t$ pions 
have experienced in average more than two reactions by the formation 
and subsequent decay of $\Delta$-resonances \cite{Bass93}, which is 
enough for at least partial thermalization of their spectrum. In 
contrast to this, high-$p_t$ pions as a probe of the hot and dense 
phase in Pb+Pb collisions at 160{\it A\/}~GeV should be handled with 
care because of the lack of rescattering for extremely energetic pions 
in the latter case. 

\begin{figure}[htp]
\centerline{\epsfysize=16cm \epsfbox{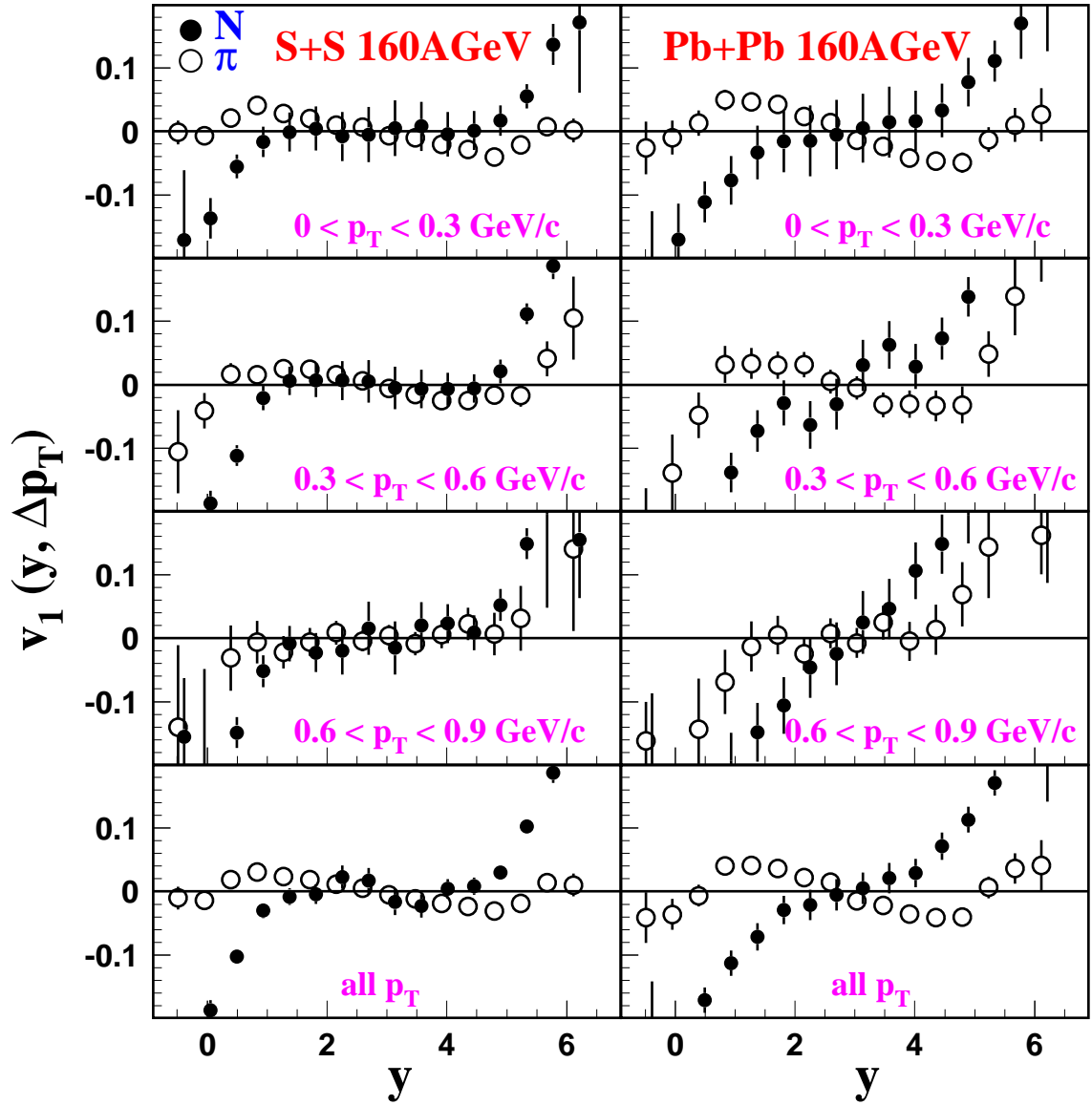}}
\caption{
Directed flow, $v_1(y,\Delta p_t)$, of pions (open circles) and
nucleons (solid circles) in different $p_t$ intervals in minimum
bias S+S (left panels) and Pb+Pb (right panels) collisions
at 160{\it A} GeV.
}
\label{fig4}
\end{figure}

The role of rescattering in the formation of directed flow is 
illustrated in Fig. \ref{fig4}. Here the directed flow as a function 
of rapidity in several $p_t$ intervals is compared in minimum bias
S+S and Pb+Pb events at 160{\it A} GeV. Lacking a sufficiently large
amount of secondary interactions per hadron, the directed flow of 
both nucleons and pions in the light S+S system varies very weakly 
with rising transverse momentum from $p_t \leq 0.3$ GeV/$c$ to 
$0.6 < p_t < 0.9$ GeV/$c$. Nevertheless, the change of the slope of
pionic directed flow from antiflow to normal flow with rising 
$p_t$ is distinctly seen for both reactions. In the heavy ion system 
the directed flow of hadrons, especially nucleons, strongly depends 
on the $p_t$ range. 
Note that, since the bulk amount of particles is produced with 
transverse momenta less than 300 MeV/$c$, the distribution, given 
by Eq.~(\ref{eq3}) integrated over the whole $p_t$ interval, is 
very close to that for $0 < p_t < 0.3$ GeV/$c$.

\subsection{$v_1(p_t)$ in rapidity intervals}
\label{subsec3c}

\begin{figure}[htp]
\centerline{\epsfysize=16cm \epsfbox{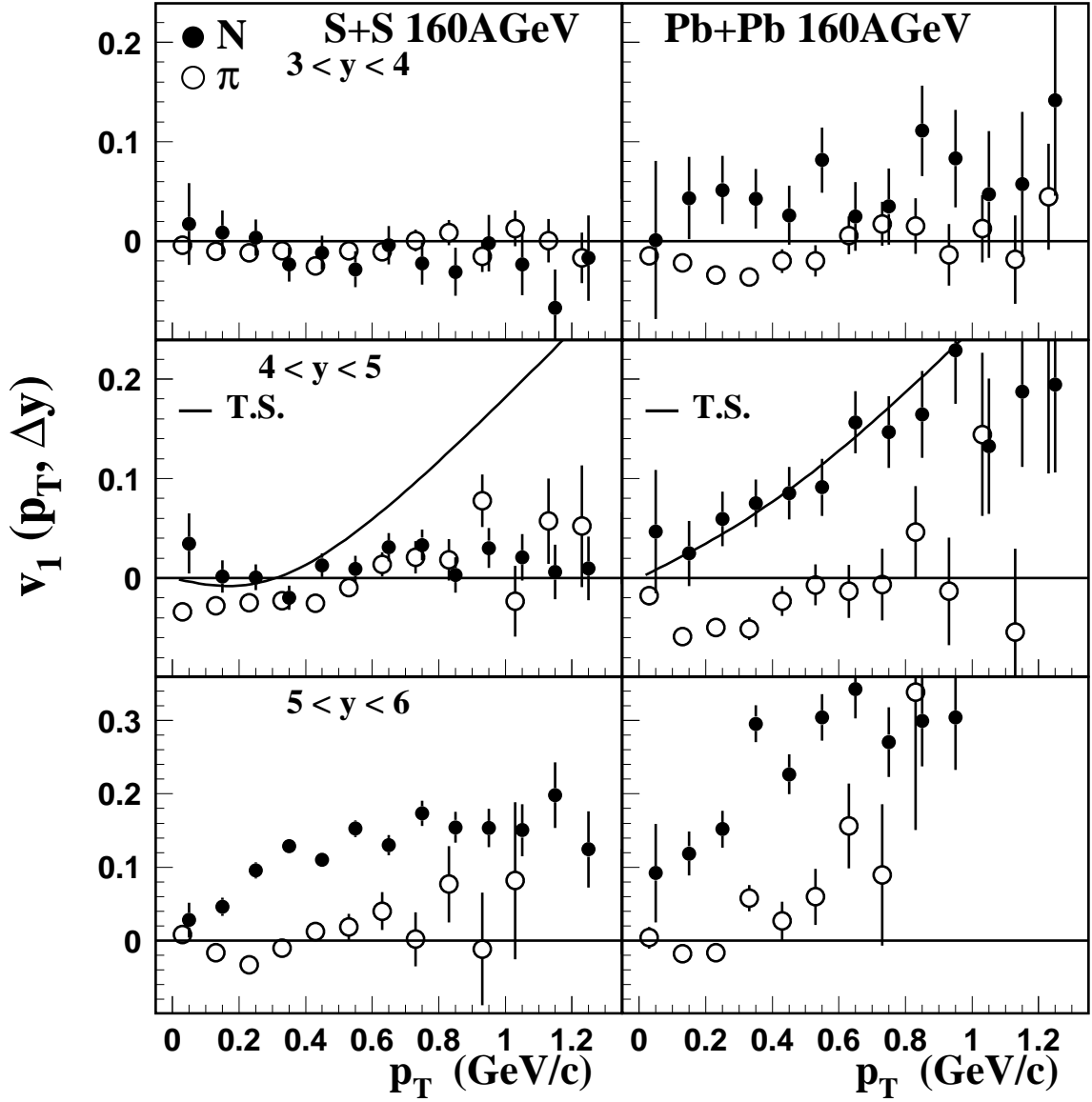}}
\caption{
Directed flow, $v_1(p_t,\Delta y)$, of pions (open circles) and
nucleons (solid circles) in different rapidity intervals in minimum
bias S+S (left panels) and Pb+Pb (right panels) collisions
at 160{\it A} GeV.
Solid curves indicate the results of the fit of
Eq.~\protect(\ref{eq4}\protect) to the protons with 
$p_t \leq 0.45$\,GeV/$c$ in the rapidity interval $4 < y < 5$. 
See text for details.
}
\label{fig5}
\end{figure}

Figure~\ref{fig5} depicts the directed flow of pions and nucleons
as a function of their transverse momentum in different rapidity
windows. Both, in S+S and Pb+Pb collisions the directed flow of 
nucleons weakly depends on $p_t$ in the midrapidity range, 
$3 < y < 4$. It starts rising with increasing $p_t$ near the
projectile/target rapidity. The directed flow of pions seems to 
increase, especially in lead-lead collisions, with rising transverse
momentum. At $p_t \leq 0.3$~GeV/$c$ the flow of pions in both 
reactions is small but negative in all three rapidity intervals. 
A similar picture is observed for low-$p_t$ nucleons with the
rapidity $y \leq 5$ in S+S collisions.
As already mentioned in the introduction, one possible 
explanation for the occurrence of antiflow at low $p_t$ is the 
collective motion of a group of particles, which are in (local)
thermal equilibrium. Using the expression for the
directed flow of non-relativistic particles emitted isotropically
by a transversely expanding thermal source \cite{Vol97}:
\beqar
\ds
\label{eq4}
v_1(p_t) &=& \frac{p_t \beta_a}{2 T} \left[ 1 -
\frac{m \beta_0}{p_t} \frac{I_1(\xi)}{I_0(\xi)} \right]\ ,\\
\xi &=& \frac{\beta_0 p_t}{T}\ , 
\label{eq5}
\eeqar
where $T$ is the temperature, $\beta_a$ is the collective velocity
along the directed flow axis, $\beta_0$ is the transverse expansion 
velocity of the source, 
and $I$ is the modified Bessel function, one can also reproduce the 
negative values of $v_1(p_t)$ in the low $p_t$ region. Note, that 
the validity of Eq.~(\ref{eq4}) for protons is restricted to
the range of $p_t \leq 0.5$ GeV/$c$. A relativistic generalisation 
of Eq.~(\ref{eq4}) to an expansion in three dimensions \cite{Vol97}
leads to an increase of about 15\% in the expansion velocity needed
to describe the low $p_t$ dip in the $v_1(p_t)$ distribution. Results 
for a fit of Eq.~(\ref{eq4}) to the low-$p_t$ part 
($p_t \leq 0.4$~GeV/$c$) of the $v_1^N(p_t)$ distribution in the
rapidity interval $4 < y < 5$ are plotted in Fig.~\ref{fig5} also. 
The fitting parameters are $\beta_a = 0.1c$, $T = 140$ MeV, and 
$\beta_0 = 0.60 (0.40)c$ for S+S (Pb+Pb) collisions. However, the 
collective flow in sulphur-sulphur system is weak \cite{Amprl91}, and 
the whole system is far from being in thermal equilibrium. Therefore, 
the plausible explanation of negative values of $v_1(p_t)$ is 
shadowing. 

It is worth noting that the flow of $\Delta$ resonances follows the
nucleon flow (see, e.g., \cite{Bass94}). Pions coming from the decays 
of $\Delta$'s behave similar to pions emitted from a moving thermal 
source \cite{e877}, i.e., the pionic flow which originates from 
$\Delta$ resonance decays should be positive at high $p_T$ and 
negative at low transverse momenta. Also, in lead-lead collisions 
local thermal equilibrium can be reached at least in the central zone 
of semicentral collisions with impact parameter $b \leq 4$~fm 
\cite{YG99,LE1,CR99}. In such events 
directed flow can be affected by a radial isotropic expansion,
the formation of long-lived $\Delta$ resonance matter, and nuclear
shadowing. In peripheral collisions the possible formation of a 
thermalized source becomes less important. Here the development of 
nucleon antiflow in the midrapidity region, as well as in the low 
$p_t$ interval, is completely determined by shadowing. 

\begin{figure}[htp]
\centerline{\epsfysize=16cm \epsfbox{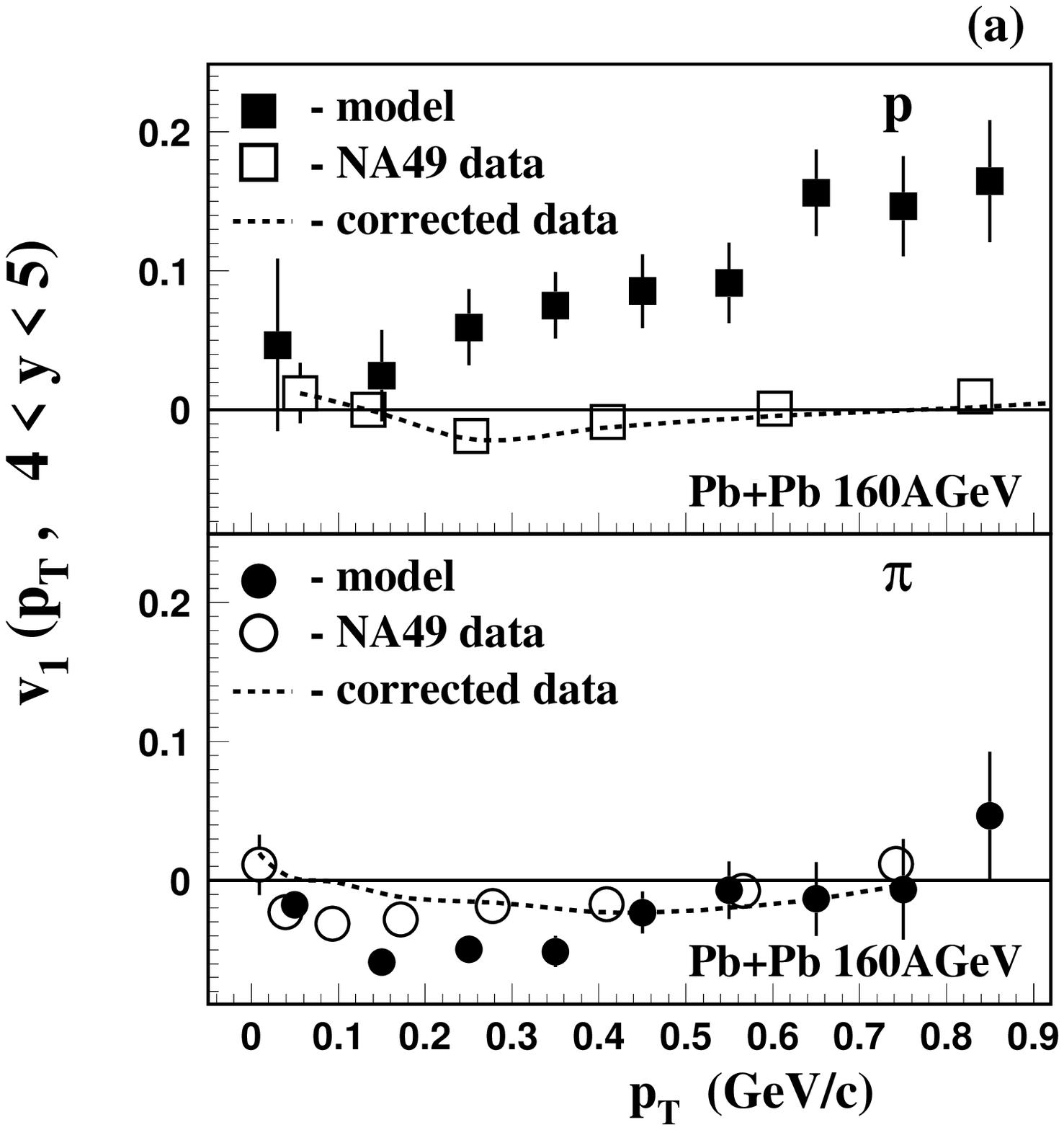}}
\caption{
(a) Directed flow, $v_1(p_t)$, of pions (open circles) and protons
(solid circles) with the rapidity $4 < y < 5$ in
Pb+Pb (upper panel) and S+S (lower panel) collisions at 
160{\it A} GeV. Open and solid stars show the experimental 
results \protect\cite{na49} on pion and proton directed flow, 
respectively. Dashed curves are the corrected data taken from
\protect\cite{BDOprc00}. 
(b) The same as Fig.\protect\ref{fig6}(a) but for protons with
$y < 0.5$ in Pb+Pb collisions. Data (open squares) are taken from 
\protect\cite{wa98b}, solid squares indicate the model predictions. 
}
\centerline{\epsfysize=16cm \epsfbox{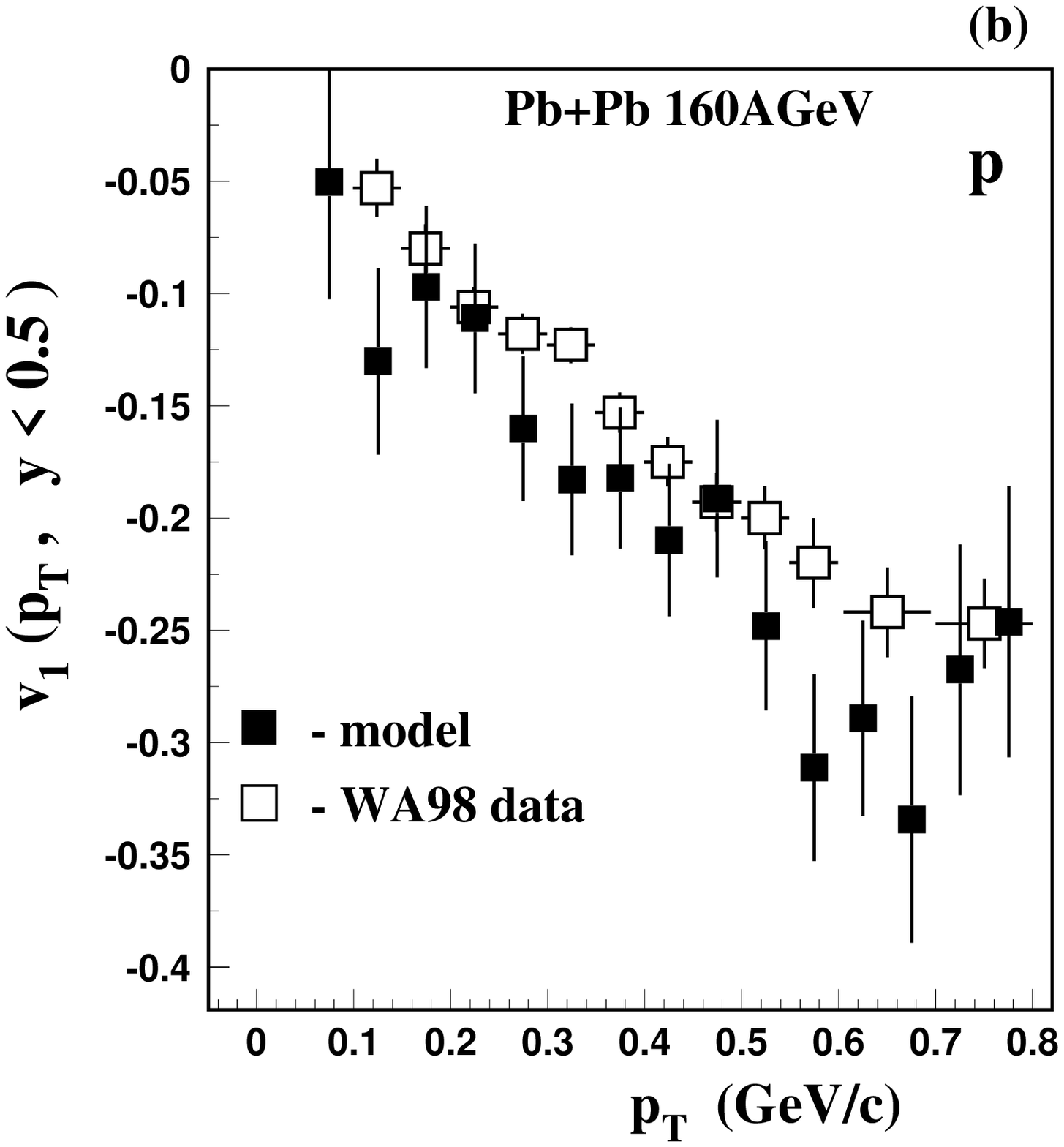}}
\label{fig6}
\end{figure}

To compare results of the microscopic calculations with
the experimental data, the directed flow of protons and pions in 
the rapidity range $4 < y < 5$ is shown separately in 
Fig. \ref{fig6}(a). One sees that the model describes the flow of
pions reasonably well, but predicts much stronger signal for protons
 at $p_t \geq 0.25$~GeV/$c$. In this context it should be noticed 
that the $p_t$ dependence measured by the NA49 Collaboration
\cite{na49} substantially deviates from the systematics seen at both
SIS \cite{andronic} and AGS \cite{e877} energies, where almost a
linear increase of the proton directed flow with $p_t$ has been
observed in Au+Au collisions. In the latter case this behaviour is
well reproduced by the RQMD simulations (see Fig.~20 of 
\cite{HWW99}). Thus, at SPS energies the QGSM
predictions are in line with the SIS/AGS flow systematics, whereas
the NA49 Collaboration observes almost a zero proton flow signal 
as a function of transverse momentum in the interval $0.1 \leq p_t 
\leq 0.9$~GeV/$c$. The present minimum bias calculations are
directly compared with the data in the corresponding rapidity
window. If this comparison is not biased by (unknown to us)
efficiency cuts (in \cite{na49} no special cuts were mentioned),
this fact, in principle,
can be taken as indication of a new dynamical feature not included
into the current version of the QGSM. It is interesting that in the
target fragmentation region, where, for instance, the formation of
the QGP is quite unlikely, the model predictions for the directed
flow $v_1^p(p_t)$ are close to the experimental data of the WA98
Collaboration \cite{wa98b}, as shown in Fig.~\ref{fig6}(b). But the
production of even a small amount of quark-gluon plasma should reduce
the strength of the low-$p_t$ pion flow too. A few explanations which 
can lead to revision of the experimental data have been proposed 
recently \cite{DBOplb00,BDOprc00}. The main argument is as follows: 
Whereas the orientation of the reaction plane is well defined in the 
microscopic calculations, its orientation in the experimental event 
has to be somehow reconstructed. Two-particle correlations in the 
azimuthal plane are usually applied for this purpose. However, as was 
discussed in detail in \cite{DBOplb00,BDOprc00}, there are several 
other sources of azimuthal particle correlations not related to the 
flow. Experimental data corrected for transverse momentum and
Hanbury-Brown-Twiss (HBT) correlations for pions, and for 
$p_t$-correlations and correlations from $\Delta$ decays for
protons \cite{BDOprc00}, are plotted onto the results of microscopic
calculations in Fig.~\ref{fig6}(a) also. We see that the directed flow 
of protons is almost not changed due to substantial mutual 
cancellations of $p_t$ correlations and $\Delta$-decay correlations, 
working in the opposite direction. The agreement between the model 
calculations and corrected data for the directed flow of low-$p_t$
pions becomes even worse. On the other hand, comparison between the 
VENUS \cite{venus} raw data sample and the data filtered through the
GEANT model of the NA49 detector seems to reveal that the sources of
non-flow correlations are insignificant \cite{na49}. This important 
problem, definitely, needs further investigations. It would be very 
interesting also to compare model predictions with the coming data 
on sulphur-sulphur collisions \cite{Rohr00}. 

\section{Conclusions}
\label{sec4}

In summary, we have for the first time performed a systematic 
study of the directed flow of nucleons and pions as a function 
of rapidity in different $p_t$ intervals at SPS energies. 
It is shown that the slope of
the directed flow of nucleons with $p_t \leq 0.6$ GeV/$c$ is positive
(normal flow) in semicentral and semiperipheral collisions, and
negative (antiflow) in very peripheral ones, where 
$b/b_{max} \geq 0.7$. At higher transverse momenta the slopes of 
both, pion and nucleon directed flow become positive. 
Our findings agree with recent experimental results which report 
a change of sign for the pion directed flow with increasing 
$p_t$ in Au+Au collisions at SIS energies \cite{Wag00}. 
There the effect was attributed solely to shadowing by the 
spectator matter. It was found also that high-$p_t$ pions 
($p_t \geq 0.4$ GeV/$c$) in heavy-ion reactions at 1{\it A\/}~GeV 
are emitted within the first 13 fm/$c$ or even earlier. 
The microscopic analysis of the particle freeze-out conditions in 
heavy-ion collisions at SPS energies shows that 
particles with maximal $p_t$ are produced essentially either in 
primary nucleon-nucleon collisions (mesons and nucleons) or in the 
early phase of the reaction within the first 17 fm/$c$ (nucleons).
In the yield of high-$p_t$ nucleons the admixture of those 
emitted within the first two fm/$c$'s is small.
This means that high-$p_t$ nucleons might be very useful to study
the nuclear equation of state at high temperatures and densities 
whereas a large amount of high-$p_t$ pions is produced even too 
early to probe this phase. 

The pion and nucleon directed flow in S+S collisions, as well as pion
directed flow in Pb+Pb collisions, indicates negative values as 
particle transverse momenta approach zero. Such behaviour can be 
explained by the interference between a radially expanded thermalized 
system and anisotropic flow \cite{Vol97}. However, since the formation 
of a rapidly expanding thermal source in peripheral heavy-ion 
collisions and in light S+S collisions is rather unlikely, the effect 
must be caused by nuclear shadowing. Hadrons emitted at small rapidity 
in the antiflow direction can propagate freely, while hadrons emitted 
in the normal flow direction still remain within the expanding 
subsystem (or core) of interacting particles. To study the interplay
between the isotropic radial flow and anisotropic directed flow one
has to subtract shadowing from the analysis of experimental data. 

The microscopic model calculations are in reasonable agreement with
most experimental data on anisotropic flow, both directed and 
elliptic, in minimum bias Pb+Pb events. The model is able to reproduce 
quantitatively the strong negative flow of pions,
$v_1^\pi (p_t, 4 < y < 5)$, at low transverse momenta, $p_t \leq
0.25$~GeV/$c$ which is seen by the NA49 Collaboration. 
Concerning the $p_t$ dependence of the proton flow the model
calculations show the same systematics as observed at SIS and AGS
energies, namely, a linear rise of the $v_1$ with $p_t$, whereas the
very weak signal at $p_t \leq 0.9$~GeV/$c$ is observed in this 
rapidity window. If the comparison with data is not
biased by unknown efficiency cuts, there is left space for the
speculation about dynamical features not incorporated into the
microscopic model. To clarify this point and to make more definite 
conclusions it would be interesting also to compare the model 
calculations with forthcoming S+S data at the same energy.  

\acknowledgments 
We are thankful to L.~Csernai, P.~Danielewicz, D.~R\"ohrich,
D.~Strottman, and H.~Wolter for the interesting discussions and 
fruitful comments.
This work was supported in part by the Bundesministerium f\"ur 
Bildung und Forschung (BMBF) under contract 06T\"U986.

\widetext

\newpage

\end{document}